\documentclass{elsarticle}

\usepackage[english]{babel}
\usepackage{booktabs} 
\usepackage{multicol} 
\usepackage{siunitx}  
\usepackage{array}    
\usepackage{amsmath}
\usepackage{graphicx}
\usepackage{caption}
\usepackage{subcaption}
\usepackage{multirow}
\usepackage{url}
\usepackage{adjustbox}
\usepackage{booktabs}
\usepackage{array}
\usepackage{float}
\usepackage{comment}
\usepackage[colorlinks=true, allcolors=blue]{hyperref}
\usepackage{xcolor}

\usepackage[letterpaper,top=2cm,bottom=2cm,left=3cm,right=3cm,marginparwidth=1.75cm]{geometry}

\title{High-Throughput Low-Cost Segmentation of Brightfield Microscopy Live Cell Images}

\begin{document}

\author[1]{Surajit Das\corref{cor1}}
\ead{mr.surajitdas@gmail.com}
\cortext[cor1]{Corresponding author}

\author[2]{Gourav Roy}
\ead{gouravroy2110@gmail.com}   

\author[1]{Pavel Zun}
\ead{pavel.zun@gmail.com} 

\address[1]{ITMO University, St. Petersburg, Russia}
\address[2]{Jadavpur University, Kolkata, India}

\begin{abstract}
Live cell culture is crucial in biomedical studies for analyzing cell properties and dynamics in vitro. This study focuses on segmenting unstained live cells imaged with bright-field microscopy. While many segmentation approaches exist for microscopic images, none consistently address the challenges of bright-field live-cell imaging with high throughput, where temporal phenotype changes, low contrast, noise, and motion-induced blur from cellular movement remain major obstacles.  

We developed a low-cost CNN-based pipeline incorporating comparative analysis of frozen encoders within a unified U-Net architecture enhanced with attention mechanisms, instance-aware systems, adaptive loss functions, hard instance retraining, dynamic learning rates, progressive mechanisms to mitigate overfitting, and an ensemble technique. The model was validated on a public dataset featuring diverse live cell variants, showing consistent competitiveness with state-of-the-art methods, achieving 93\% test accuracy and an average F1-score of 89\% ($\pm$0.07) on low-contrast, noisy, and blurry images.  

Notably, the model was trained primarily on bright-field images with limited exposure to phase-contrast microscopy ($<$20\%), yet it generalized effectively to the phase-contrast LIVECell dataset, demonstrating modality, robustness and strong performance. This highlights its potential for real-world laboratory deployment across imaging conditions.  

The model requires minimal compute power and is adaptable using basic deep learning setups such as Google Colab, making it practical for training on other cell variants. Our pipeline outperforms existing methods in robustness and precision for bright-field microscopy segmentation. The code and dataset are available for reproducibility \footnote{This paper is under review. The full dataset and code used in this study may be provided upon reasonable request to the corresponding author}.

\end{abstract}

\begin{keyword}
    Live cell segmentation \sep Microscopy \sep Multi-cell line validation \sep Colab deployment \sep Computational bioimaging
\end{keyword}

\maketitle

\section{Introduction}

Quantification and segmentation of cells are foundational tasks in biological research, critical for analyzing cell morphology, behavior, and function \cite{zhu2017extended,ferreira2024classification}. Among various imaging techniques — bright-field, phase-contrast, fluorescence, electron, and confocal microscopy — bright-field microscopy remains a widely used modality due to its simplicity, cost-effectiveness, and ability to image unstained, live cells without cytotoxic dyes or complex preparations \cite{chen2011optical,mualla2013automatic,lulevich2009cell}.

Despite its accessibility, bright-field microscopy poses significant segmentation challenges. Live cell images often suffer from low contrast, uneven illumination, overlapping structures, and noise arising from culture medium debris, gas bubbles, and cell movement during imaging \cite{cheng2014neurosphere,chen2015bright}. These issues complicate conventional segmentation and demand more robust, adaptive solutions.

Manual segmentation — still used in many labs — is time-consuming and inconsistent, especially with large datasets \cite{wiggins2025exploring}. Automation using deep learning (DL), particularly Convolutional Neural Networks (CNNs), has emerged as a powerful alternative for image segmentation across modalities, including microscopy, medical imaging, and remote sensing \cite{khalifa2023deep,minaee2021image}. Architectures like U-Net \cite{ronneberger2015u}, ResNet \cite{chen2023application}, and LinkNet \cite{rajesh2022prostate} have been widely adopted due to their ability to learn multi-scale spatial features and delineate fine structures.

However, domain-specific constraints in bright-field microscopy — such as subtle textures, low signal-to-noise ratios, and minimal contrast — can significantly degrade performance unless models are carefully optimized for this context \cite{wu2014investigation,bradbury2010spectral}. Generic CNNs trained on natural or stained image datasets often fail to generalize without adaptation.

In this study, we develop a robust CNN-based segmentation pipeline specifically tailored to bright-field microscopy of unstained live cells. Our architecture incorporates frozen encoder backbones, attention mechanisms, instance-aware processing, adaptive loss functions, hard-instance retraining, and ensemble learning—strategies designed to counteract low-contrast and noisy imaging conditions. We use manual ground truth masks due to the poor performance of automated tools like Cellpose and StarDist on these difficult images.

Importantly, while our training set was composed primarily of bright-field images, it included a small proportion $(<20\%)$ of phase-contrast images, allowing us to evaluate the model's robustness across imaging modalities. The model generalized effectively to phase-contrast images in the LIVECell dataset, demonstrating its adaptability and cross-modality potential.

This paper presents our pipeline and results, validated on diverse cell lines (e.g., A549, C2C12, A172), and discusses its implications for high-throughput, low-cost segmentation in biomedical imaging and regenerative medicine research.

\section{Related Work}
 CNN-oriented methodologies for cellular segmentation and classification, along with a variety of architectures, pre-processing steps, and evaluative metrics have been explored by numerous experiments. Based on thematic categorizations, the researches can be categorized into three primary groups, namely, 1) CNN-based segmentation in bright-field microscopy, 2) cell classification and morphological analysis, and 3) advanced DL-based segmentation techniques. This section presents the recent studies with the accent on identifying methodological advancements and achievements along with the current research gaps, and simultaneously articulates the significance of the current study pertaining to myoblast cell segmentation and morphological analysis.

\subsubsection*{\textbf{CNN-based segmentation in bright-field microscopy:}} Incorporating a variety of supervised learning models, frameworks and pre-processing steps, researchers have established the superior performance of CNN in terms of accuracy in the segmentation problems in the field of bright-field microscopy data. In one specific work, ScoreCAM-U-Net \cite{ali2022artseg} combined artifact removal with a weakly supervised CNN-based segmentation technique to improve the quality of microscope pictures. Even though this approach was successful in producing significant segmentation, it had trouble differentiating between different kinds of artifacts and required more advancements to improve generalisation across a range of datasets. \\

Another study evaluated residual attention U-Net architectures for semantic segmentation of living HeLa cells in bright-field transmitted light microscopy, achieving good performance for delineating individual live cells in challenging, label-free imaging scenarios \cite{ghaznavi2022105805}. Moreover, investigations employing U-Net-based architectures have indicated superior segmentation of entire cells through the utilization of cytoplasmic markers instead of nuclear stains \cite{al2018deep}. Correspondingly, techniques involving edge detection and morphological operations have been employed for bright-field segmentation \cite{vcepa2018segmentation}; however, these methodologies frequently encounter difficulties with indistinct cellular boundaries (having the same colour as background) and necessitate extensive manual
parameter optimization.

\subsubsection*{\textbf{Cell classification and morphological analysis:}} Many investigations have applied CNNs in the domain of cell classification. A specific investigation, which is concentrated on bright-field microscopic images, leveraged CNN models for the classification of unstained cells \cite{ferreira2024classification}, and achieved the accuracy of 93\%.  For instance segmentation, some methodologies, like Gene-SegNet \cite{wang2023genesegnet} and Mesmer \cite{greenwald2022whole}, have been proposed. The idea is to integrate deep learning architectures for cell segmentation with advanced feature extraction mechanisms. Gene-SegNet synthesized imaging and gene expression data to enhance segmentation capabilities, whereas Mesmer attained human-level segmentation
accuracy by utilizing extensive annotated datasets. Inspite of producing high segmentation efficacy, they place considerable emphasis on dataset-driven training and generalization, with limited focus on the morphological analysis of individual segmented cells.\\

Another important research introduces NeuSomatic \cite{sahraeian2019deep}, which considers the CNN model for the purpose of somatic mutations detection. TissueNet \cite{greenwald2022whole}, well-known for having a large-scale dataset which is designed to train segmentation models, could work proficiently in whole-cell identification. While these investigations have made significant contributions to deep learning-driven biomedical imaging, they have not specifically addressed the segmentation of live myoblast cells within the domain of bright-field microscopy.

\subsubsection*{\textbf{Advanced DL-based segmentation techniques:}} Researchers have thought of introducing some unsupervised learning methods (deep learning methodologies) for segmentation tasks without using manual annotation\cite{Din2021.05.17.444529}. There are also the examples of semi-supervised learning \cite{lam2025ssl} approach. A modified U-Net architecture with the facility of marker-controlled segmentation has been proposed for both bright-field and fluorescence microscopy images. It is found that the model has substantially enhancing efficiency while manual annotation has been minimized. It is important to mention that utilization of recursive training strategies in automated segmentation pipelines have exhibited high intersection-over-union (IoU) scores. Hybrid deep learning approaches have also been explored, amalgamating traditional segmentation techniques, such as watershed algorithms, with CNNs to achieve improved accuracy \cite{Fotos2023Deep}. Despite their successes, these models frequently require extensive training datasets. Also, often they become computationally expensive architectures which is not always pragmatic in respect of usability, thereby posing challenges for real-time processing.

Although deep learning-driven segmentation methodologies have attained good and remarkable results when applied across the range of diverse microscopy images, several problems persist as stated below:
\begin{itemize}
\item Lack of customized CNN architectures for live myoblast cells: contemporary segmentation techniques predominantly focus on generic cellular classifications, with a lacking of specific focus on the segmentation of myoblast cells within bright-field microscopic field.
\item Inadequate post-segmentation morphometric assessment: in most cases, it is observed that the investigators prioritize
segmentation precision but fail to undergo the analysis of the cells by applying morphological metrics such as convexity, circularity, aspect ratio, area, etc.
\item Lack of comparative morphological analysis: there exist very few studies which put forward comparison among the morphological statistics of segmented cells, and hence, the scope of posit any significant augments is limited which affects the comprehension of cellular differentiation and behavior.

\end{itemize}

\subsubsection*{\textbf{New SOTA Releases:}}
Recent advancements in segmentation methodologies have introduced two notable approaches that warrant examination in the context of bright-field myoblast analysis:

\paragraph{\textbf{Self-Supervised Learning (SSL) Approaches}}
The work by \cite{lam2025ssl} represents a significant shift toward annotation-free segmentation through optical flow-based pseudo-labeling. While achieving $F_1$ scores of 0.77--0.88 on fluorescence images, our evaluation revealed critical limitations for bright-field applications: (1) processing times of 50--60 seconds per image due to iterative optical flow computation, and (2) complete failure on 60\% of low-contrast myoblast samples where texture features proved unreliable. These constraints are particularly problematic for longitudinal studies requiring both speed and consistency across imaging sessions. The error message ``Either no cells found or all cells are touching the border" typically occurs in image analysis or cell segmentation tasks when the algorithm fails to properly identify cells or detects cells that are too close to the image edges. 

\paragraph{\textbf{Cellpose Evolution}}
The latest version, \textbf{Cellpose-SAM}~\cite{pachitariu2025cellposesam}, introduced by Pachitariu \textit{et al.} (2025), integrates Segment Anything Model (SAM) components by combining ViT-L encoders with flow-based decoding, achieving a reported 15\% IoU improvement on phase-contrast data over earlier Cellpose releases. While this framework claims broad ``segment anything'' capability, our experiments reveal that this generalization does not extend to unstained bright-field live-cell microscopy, where poor performance is observed (Tab~\ref{tab:detailed_performance}). We attribute this degradation to a domain shift from natural image pretraining. Furthermore, the model's higher hardware requirements (minimum 16~GB VRAM) hinder its adoption in typical laboratory workstations. These findings highlight a gap between claimed universality and domain-specific performance, underscoring the trade-off between architectural complexity and practical deployment in biological labs and clinical microscopy.

The contribution and value of the present research lie with mainly addressing the following research gaps. In contrast to existing methodologies that emphasize general cell types, this investigation implements CNN architectures explicitly for the segmentation of living myoblast cells in bright-field microscopy, thereby optimizing performance for this specific cell type and imaging conditions. Additionally the study approaches to a holistic morphometric analysis. Following segmentation, this research quantifies essential morphological characteristics such as convexity, circularity, aspect ratio, and area, thereby furnishing more profound insights into the morphology of myoblast cells and showing the procedure of formally assessing their phenotype.

\section{Methodology:}

\subsection{Setting Low-Compute Environment:}
Training (max 6.5 hours) was performed on \textbf{Google Colab} using a Linux system (kernel 6.1.123+), with an Intel Xeon CPU (4 cores, 8 threads), 12~GB RAM, and an NVIDIA Tesla T4 GPU (16~GB VRAM). The pipeline was built in TensorFlow~2.18.0 with CUDA, using Python~3.10. Core libraries included OpenCV~4.11.0 (image preprocessing), scikit-image (morphological ops), NumPy~2.0.2, and Pandas~2.2.2. Annotations were generated via CVAT.

\subsection{Acquiring Dataset:}
The dataset consists of 697 (256x256) bright field microscopy images and 160 phase contrast microscopy images of unstained live cells in culture medium. Therefore, the total number of instances is 857. The dataset is obtained by our research group.\\ The dataset is fairly complex in nature for any sort of automated end-to-end analysis for the following reasons.\\

\begin{figure}[H]
  \centering  \includegraphics[width=1\textwidth]{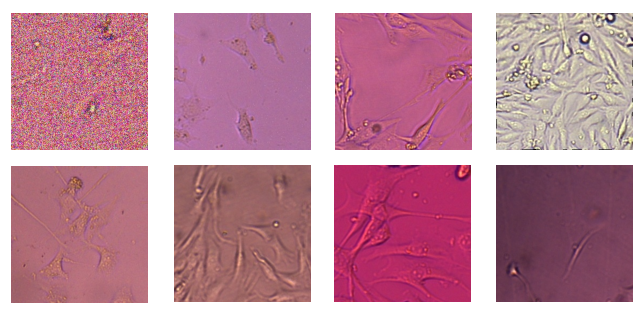}
  \caption{Instances used in the model pipelines}
  \label{cell-instance}
\end{figure}

First of all, it contains overlapping cellular structures. Secondly, the morphology of a cell usually changes when it is ready to divide, or when it moves along the substrate, or when it is ready to differentiate into a particular tissue. Accordingly, when a cell is ready to divide, it becomes slightly larger than the existing ones, and its nucleus also increases. The cell can become more rounded. When a cell moves along the substrate, it can stretch along some axis. And when a cell is about to differentiate, then depending on the type, it can become either a star-like cell (then it will turn into bone tissue), or an elongated cell (then it will turn into muscle).
In some case, all the changes in morphology are within the normal range and they show that the cell is simply moving along the substrate.

Next complexity lies with the noise. The most common noise that appears in the picture can be caused by protein molecules that are part of the culture medium, as well as by protein molecules adsorbed to the substrate next to the cell. These molecules are usually produced by cells during their life processes. Noise can also be caused by shadows of oxygen bubbles that float in the culture medium. Sometimes water vapor can condense on the upper lid of the Petri dish, which can also cast a shadow and create noise.

Lastly, the image plane itself contains some noise due to the imperfect calibrations of the imaging system, which cause illumination non-uniformity, optical aberrations, improper condenser alignment, diffraction effects, etc. Fig~\ref{cell-instance} demonstrates some samples of data considered for training analysis.

\subsection{Data Pre-Processing:}
\subsubsection{Masking:} 
The masks of the data were generated manually by the subject matter experts with the help of CVAT (Computer Vision Annotation Tool). During masking, the following points are taken into account. $(i)$ A cell is primarily identified by its nucleolus and shape around the nucleolus. Sometimes round light dots are visible without any nucleoli inside the objects, which are most likely not nuclei  but just debris. $(ii)$ The morphology of a cell usually changes when it is ready to divide, or when it moves along the substrate, or when it is ready to differentiate into a particular tissue. Accordingly, when a cell is ready to divide, it becomes slightly larger than the quiescent ones, and its nucleus also increases in size. $(iii)$ The cell can become more rounded. The cell should not exit the average statistical measures.

\begin{figure}[H]  
  \centering  \includegraphics[width=1\textwidth]{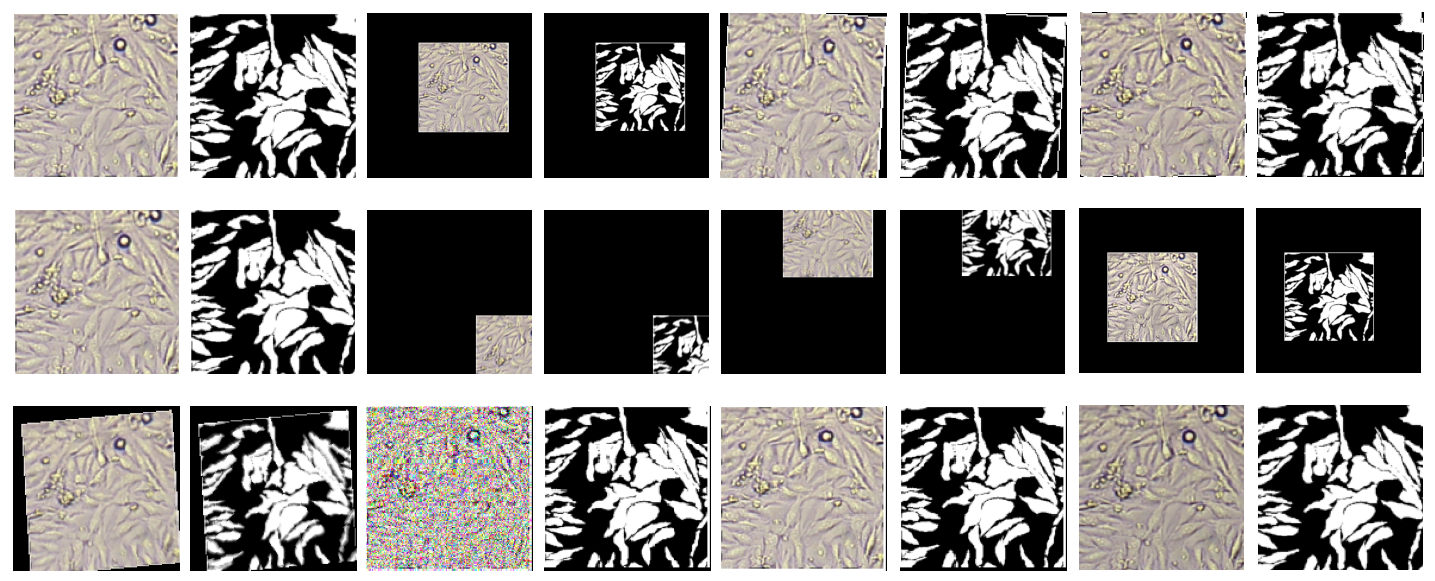}
  \caption{Different augmentation results of an instance along with its mask. Top row (left to right): Original Image, Original Mask, Original+PadAndCrop, Original+PadAndCrop (mask), Original+DivisionShift, Original+DivisionShift (mask). Middle row: Original+MotionCrop, Original+MotionCrop (mask), MotionBlur+PadAndCrop, MotionBlur+PadAndCrop (mask), Defocus+PadAndCrop, Defocus+PadAndCrop (mask). Bottom row: ZoomBlur+CrowdingAffine, ZoomBlur+CrowdingAffine (mask), GaussianNoise+GridDistort, GaussianNoise+GridDistort (mask), ISONoise+GridDistort, ISONoise+GridDistort (mask).}
\end{figure}

\subsubsection{Data Augmentation}
\label{subsec:augmentation}

We implemented an extensive augmentation pipeline using the \texttt{albumentations} library (v1.4.3). Each image--mask pair was expanded into 21 variants, yielding a total of 17997 training samples. The augmentation strategy consisted of two sequential transformation stages: a primary photometric transformation followed by a secondary geometric transformation.

Primary transformations were applied to the images only, except where necessary to preserve label fidelity. These included synthetic optical effects such as motion blur (with kernel sizes ranging from 3 to 7 pixels) and defocus (radius 1--3 pixels), as well as noise models like Gaussian noise and ISO noise to simulate sensor artifacts. Color and contrast alterations such as hue--saturation--value shifts, RGB channel shuffling, brightness--contrast jittering, gamma correction, and CLAHE were introduced to reflect common sources of illumination and staining variability in brightfield microscopy. Some transformations, such as grayscale conversion and channel inversion, were used to enforce invariance to color information.

Following the photometric step, a random geometric transformation was applied to both the image and the corresponding binary mask. These included elastic deformations with moderate $\alpha$ and $\sigma$ values to mimic cellular elasticity, grid distortions to simulate spatial warping, and affine transforms incorporating scaling, translation, rotation, and shearing to emulate mitotic shape changes. We also employed patch-based augmentations such as random resized cropping and padding followed by random cropping, which encouraged the model to learn from diverse spatial contexts. Additional augmentations included horizontal and vertical flipping, 90-degree rotations, and transposition to enhance rotational and reflectional invariance.

All geometric transformations were applied using nearest-neighbor interpolation to ensure that the masks remained binary and topologically consistent. Additionally, mask binarization was enforced using a fixed threshold of 127 on 8-bit grayscale images. Each augmented sample was composed of one primary transformation (or left unaltered for baseline copies) followed by one randomly selected secondary transformation. This dual-stage augmentation pipeline was specifically designed to simulate the types of variation commonly observed in brightfield microscopy, including optical artifacts, biological heterogeneity, and staining inconsistencies. As a result, the augmented dataset contributed significantly to the model’s ability to generalize, helping it achieve a test accuracy of 93\% without overfitting to the small original training set.

\begin{figure}[H]  
  \centering  \includegraphics[width=1\textwidth]{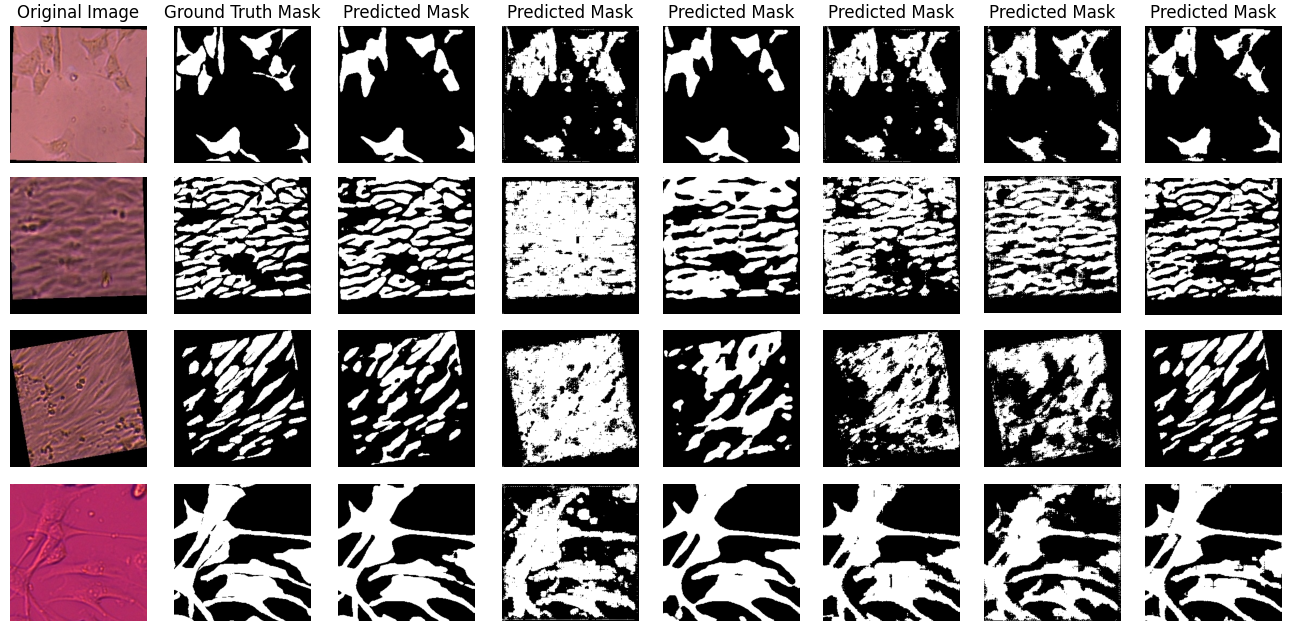}
  \caption{From left: Image, Ground truth, Segmentation result produced by Densenet, Efficientnet, Inception, Mobilenet, Resnet, Vgg16 respectively.}
\end{figure}

\subsubsection{Test-Train Split:} The dataset is split into two subsets namely, Train and Validation with the proportion 80:20.

\subsection{Model Architecture \& Training }
\subsubsection{A Pre-Selection Walkthrough for U-NET Backbone:}
We implemented six U-Net variants with frozen encoder backbones (DenseNet121, InceptionV3, VGG16, MobileNetV2, ResNet50, and EfficientNetB0) under identical training protocols for microscopy image segmentation.

All models used $\SI{256}{\times\SI{256}{}}$ patches extracted from $\SI{1536}{\times\SI{2048}{}}$ source images, trained for up to 45 epochs (batch size=16) with a combined Dice and weighted binary cross-entropy loss (10:1 class ratio), Adam optimization (initial $LR=\num{1e-4}$ with 0.9 decay every $\num{10e3}$ steps), and consistent data augmentation. The VGG16 variant completed all epochs without triggering early stopping (patience=5), while other architectures showed varied convergence patterns. Performance was tracked using standard metrics (FN/FP/TN/TP, accuracy, F1, IoU, precision, recall) with full TensorBoard logging under controlled hardware/software conditions.

\begin{figure}[H]
\centering
\includegraphics[width=1\linewidth]{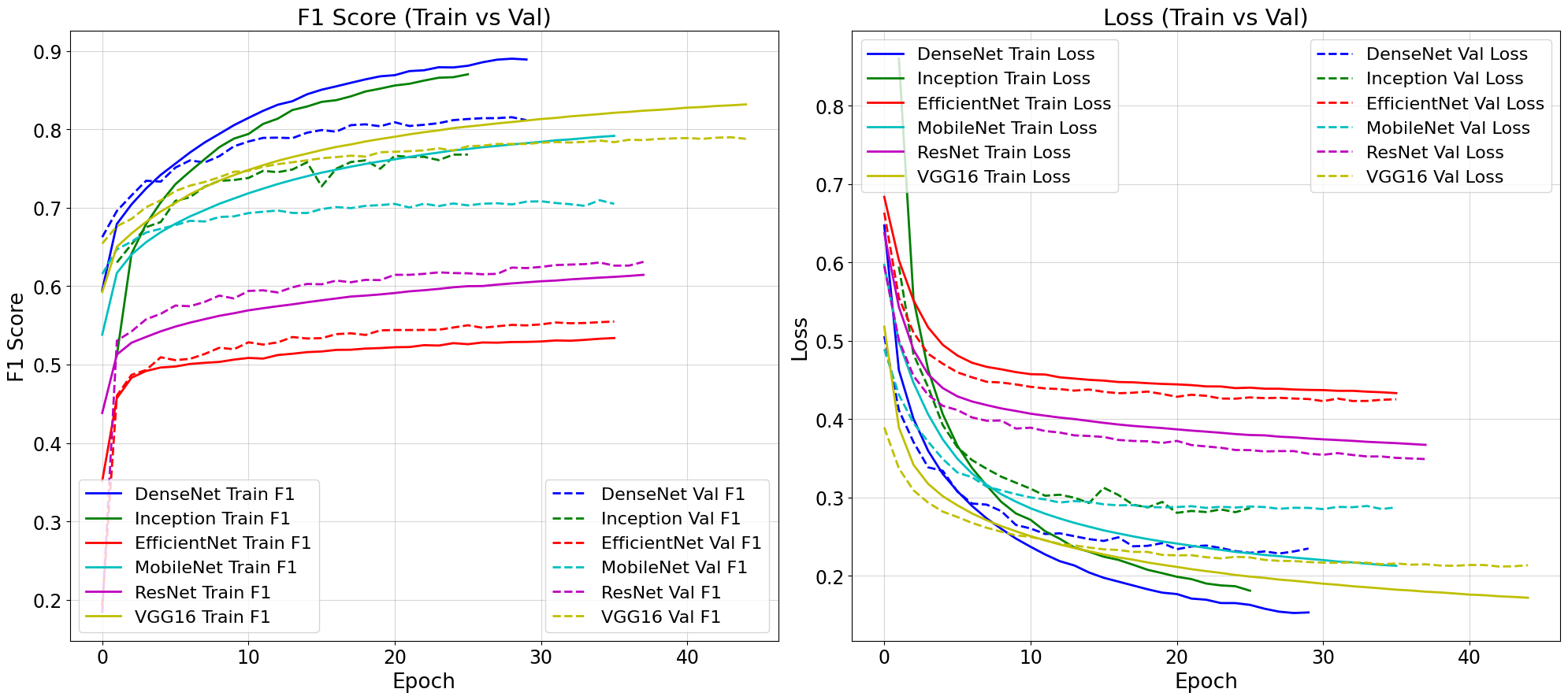}
\caption{(Left) Validation F1-score comparison across all models. DenseNet achieves the highest score, followed closely by Inception and VGG16. (Right) Validation loss comparison across all models. DenseNet and MobileNet exhibit the lowest validation losses, suggesting better optimization and reduced overfitting.}
\label{fig:val_f1_score}
\end{figure}

\begin{table}[H]
\centering
\footnotesize
\caption{Comprehensive Model Comparison} 
\label{tab:performance}

\begin{adjustbox}{width=0.90\textwidth}
\begin{tabular}{@{}lcccccc@{}}
\toprule
\textbf{} & \textbf{DenseNet121} & \textbf{InceptionV3} & \textbf{Vgg16} & \textbf{MobileNetV2} & \textbf{ResNet50} & \textbf{EfficientNetB0} \\
\midrule

\multicolumn{7}{@{}l}{\textbf{Architecture Features}} \\
\multirow[t]{3}{*}{\rotatebox[origin=c]{90}{\parbox[t]{1.2cm}{\centering\footnotesize Key Feat.}}}
& \begin{tabular}[c]{@{}l@{}}\textbullet~Dense blocks\\\textbullet~Feature reuse\\\textbullet~Concat\end{tabular} 
& \begin{tabular}[c]{@{}l@{}}\textbullet~Parallel convs\\\textbullet~Factorized\\\textbullet~Auxiliary\end{tabular} 
& \begin{tabular}[c]{@{}l@{}}\textbullet~Seq 3×3\\\textbullet~Max-pool\\\textbullet~Homog\end{tabular} 
& \begin{tabular}[c]{@{}l@{}}\textbullet~Inv res\\\textbullet~DW sep\\\textbullet~Lin bott\end{tabular} 
& \begin{tabular}[c]{@{}l@{}}\textbullet~Res blocks\\\textbullet~Skip conn\\\textbullet~Bottleneck\end{tabular} 
& \begin{tabular}[c]{@{}l@{}}\textbullet~MBConv\\\textbullet~Swish\\\textbullet~Comp scale\end{tabular} \\

\midrule

\multicolumn{7}{@{}l}{\textbf{Performance Metrics}} \\
$L_{\text{final}}$           & 0.1535            & 0.1812             & 0.1722         & 0.2130             & 0.3674          & 0.4334               \\
Gradient                      & 0.0282            & 0.0315             & 0.0184         & 0.0219             & 0.0166          & 0.0157               \\
$L_{\text{train}}$           & 0.1535            & 0.1812             & 0.1722         & 0.2130             & 0.3674          & 0.4334               \\
$L_{\text{val}}$             & 0.2351            & 0.2866             & 0.2765         & 0.2876             & 0.3492          & 0.4255               \\
$\Delta L$                  & 0.0816            & 0.1054             & 0.1043         & 0.0746             & $-$0.0182       & $-$0.0079            \\
Epochs $E$                  & 30                & 26                 & 45             & 36                 & 38              & 36                   \\
$CR$                        & 0.0623            & 0.0606             & 0.0431         & 0.0416             & 0.0259          & 0.0225               \\
$F1_{\text{train}}$          & 0.9603            & 0.9542             & 0.9436         & 0.9365             & 0.8728          & 0.8520               \\
$F1_{\text{val}}$            & 0.9356            & 0.9181             & 0.9064         & 0.9079             & 0.8749          & 0.8555               \\
$\Delta F1$                 & 0.0247            & 0.0361             & 0.0372         & 0.0286             & $-$0.0021       & $-$0.0035            \\
ORI                         & 0.0797            & 0.1017             & 0.1005         & 0.0725             & $-$0.0181       & $-$0.0078            \\
$\sigma_{\text{F1-val}}$     & 0.0068            & 0.0092             & 0.0111         & 0.0074             & 0.0126          & 0.0133               \\
$\sigma_{\text{Loss-val}}$   & 0.0075            & 0.0097             & 0.0105         & 0.0082             & 0.0142          & 0.0128               \\
$S_{\text{F1}}$              & 147.06            & 108.70             & 90.09          & 135.14             & 79.37           & 75.19                \\
$S_{\text{Loss}}$            & 133.33            & 103.09             & 95.24          & 121.95             & 70.42           & 78.13                \\
$\mathcal{O}(m)$            & 0.087             & 0.118              & 0.052          & 0.109              & $-$0.027        & $-$0.036             \\

\midrule

\multicolumn{7}{@{}l}{\textbf{Parameters (M)}} \\
Trainable & 15.65 & 19.86 & 5.53 & 1.39 & 31.05 & 2.04 \\
Total & 22.70 & 41.66 & 20.25 & 3.65 & 54.64 & 6.09 \\
\bottomrule
\end{tabular}
\end{adjustbox}

\vspace{0.1cm}
\footnotesize
Key: $CR$=Convergence Rate, $S$=Stability Index, $\mathcal{O}(m)$=Overfitting Coefficient
\end{table}

Our comparative analysis examined multiple perspectives: convergence dynamics through epoch-wise metrics, generalization gaps between training/validation performance, architectural differences via quantitative benchmarks (Table~\ref{tab:performance}), error pattern distributions, and model stability across training runs. This holistic evaluation identified VGG16 as the most robust backbone, demonstrating superior segmentation performance and training stability for our live cell culture images while maintaining computational efficiency through frozen encoder weights and optimized decoder blocks.

\subsubsection{Architecture of MODEL-1}
To address the segmentation task with both high-level semantic understanding and fine-grained spatial accuracy, we developed a hybrid encoder--decoder architecture referred to as \textbf{MODEL-1}. This model synergistically combines a pre-trained DenseNet-121 encoder with a custom-designed U-Net--style decoder, augmented with progressive dropout, spatial alignment, and regularization techniques.

\paragraph{\textbf{Encoder}} The encoder is based on DenseNet-121, a densely connected convolutional neural network pre-trained on ImageNet. This backbone is employed to extract hierarchical feature representations at multiple spatial resolutions. All encoder layers are frozen during training to retain the generalization capacity of the pre-trained features. Feature maps are extracted from intermediate layers and used as skip connections from the following layers: \texttt{conv1 relu} at 128~$\times$~128 resolution, \texttt{pool2 relu} at 64~$\times$~64, \texttt{pool3 relu} at 32~$\times$~32, \texttt{pool4 relu} at 16~$\times$~16, and \texttt{relu} at 8~$\times$~8, which is used as the bridge between the encoder and decoder.

\paragraph{\textbf{Decoder}} The decoder reconstructs the segmentation map via a series of upsampling blocks. Each block consists of a transposed convolution (\texttt{Conv2DTranspose}) for spatial upsampling, followed by batch normalization and ReLU activation, and then a dropout layer whose rate increases progressively in shallower layers. Bilinear resizing is applied to feature maps for alignment, after which they are concatenated with the corresponding encoder skip connections. Following the decoder path, additional convolutional layers are used for final refinement. A \texttt{Conv2DTranspose} layer upsamples the feature map to 256~$\times$~256, and a final 1~$\times$~1 convolution layer with a sigmoid activation produces the binary segmentation mask. After the decoder path, additional convolutional layers perform final refinement. Specifically, a \texttt{Conv2DTranspose} layer upsamples the output to a spatial resolution of $256 \times 256$, followed by a $1\times1$ convolution layer with sigmoid activation to generate the final binary segmentation mask.

\paragraph{\textbf{Regularization}}
All convolutional layers are L2-regularized with a weight decay coefficient $\lambda = 10^{-4}$. Progressive dropout is applied at rates ranging from 0.25 to 0.5, depending on depth. 

\paragraph{\textbf{Output}}
The final output is a single-channel segmentation map with values in the range $[0, 1]$. 

The model has 22.7 million total parameters (15.7M trainable) and custom Lambda layers for intermediate tensor operations.

\subsubsection{Architecture of MODEL-2}
The proposed architecture is a modified U-Net that integrates an ImageNet-pretrained VGG16 encoder with a lightweight attention-guided decoder, optimized for high-resolution biomedical image segmentation. To enhance training efficiency and numerical stability on modern hardware, mixed-precision training~\cite{micikevicius2018mixed} was employed using TensorFlow's automatic policy casting.

\paragraph{\textbf{Encoder}}
We adopted the convolutional backbone of VGG16~\cite{simonyan2014very}, pretrained on ImageNet and truncated at the final convolutional block (\texttt{block5\_conv3}). This encoder comprises five convolutional stages, each followed by max pooling. From the intermediate layers (\texttt{block1\_conv2}, \texttt{block2\_conv2}, \texttt{block3\_conv3}, and \texttt{block4\_conv3}), the feature maps serve as skip connections to the decoder. All convolutional layers in the encoder were frozen to retain pretrained semantic priors, although selective fine-tuning can be enabled.

\paragraph{\textbf{Attention-Enhanced Decoder}}
The decoder consists of a series of upsampling and convolutional blocks that progressively reconstruct the spatial resolution. At each decoding stage, the upsampled feature maps are concatenated with encoder feature maps modulated via an attention gate mechanism~\cite{oktay2018attention}. The attention gate computes an additive attention signal by aligning encoder features $x$ with decoder context $g$ through intermediate transformations:
\begin{equation}
    \psi = \sigma\left( \mathrm{Conv}_{1 \times 1}(\mathrm{ReLU}(\mathrm{Conv}_{1 \times 1}(x) + \mathrm{Conv}_{1 \times 1}(g))) \right),
\end{equation}
where $\sigma$ denotes the sigmoid activation. The output attention map $\psi$ is applied via element-wise multiplication to suppress irrelevant encoder activations.

Each decoder block follows this gated fusion with two $3 \times 3$ convolutional layers, each followed by batch normalization and ReLU activation, optionally including dropout ($p=0.1$). The number of filters decreases at each stage to reduce memory consumption, making the model deployment-friendly on constrained hardware (e.g., NVIDIA T4 GPUs).

\paragraph{\textbf{Output Layer}}
The final decoder output is passed through a $1 \times 1$ convolution with sigmoid activation to produce the segmentation mask:
\begin{equation}
    \hat{Y} = \sigma(\mathrm{Conv}_{1 \times 1}(f_d)),
\end{equation}
where $f_d$ denotes the final decoder feature map.

\paragraph{\textbf{Model Summary}}
The complete IAUNet model comprises $\sim$18.8 million parameters, of which 4.09 million are trainable due to the frozen VGG16 encoder. The architecture was implemented using TensorFlow 2.x and trained using the \texttt{mixed\_float16} precision policy. The model maintains a balance between segmentation accuracy and computational efficiency, leveraging both pretrained semantic knowledge and task-specific spatial adaptivity via attention.

\subsubsection{Architecture of MODEL-3:}
\textbf{MODEL-3} is a hybrid encoder–decoder architecture that integrates the U-Net design with pre-trained backbone encoders (VGG16), augmented by attention gating and a novel residual-style \textit{Instance Activation (IA)} module. The model is designed for dense prediction tasks such as semantic segmentation, with emphasis on robust feature recovery, contextual filtering, and stable gradient propagation.

\paragraph{\textbf{Encoder (Backbone)}}

The encoder utilizes a pre-trained ImageNet backbone (VGG16), truncated at five hierarchical feature extraction stages. Feature maps are extracted from the outputs of the max-pooling layers at the end of each convolutional block, ranging from \texttt{block1\_pool} through \texttt{block5\_pool}. All encoder weights are frozen during training to preserve pre-trained representations and mitigate overfitting, particularly in low-data regimes.

\paragraph{\textbf{Decoder}} The decoder follows a symmetric architecture with transposed convolution layers for upsampling, followed by a residual convolutional block at each stage. Each decoder block includes two $3{\times}3$ convolution layers with ReLU activation and batch normalization, an Instance Activation (IA) module for local feature recalibration, and spatial dropout layers with rates increasing from 0.2 to 0.4 as the network depth increases. Skip connections are incorporated at each stage, concatenating the encoder features with the decoder outputs at the corresponding resolution.

\paragraph{\textbf{Attention Gates}}

Attention Gates (AGs) are integrated into each skip connection to refine the fusion between encoder and decoder features. These gates compute an additive attention signal between the encoder features and a gating signal from the decoder, producing a spatial attention mask that suppresses irrelevant activations while enhancing salient structures. This mechanism improves the network’s focus on target regions without adding computational overhead.

\paragraph{\textbf{Instance Activation Module}}

The Instance Activation (IA) module is a lightweight residual block designed to enhance local activations. It comprises a $1{\times}1$ convolution layer followed by batch normalization and ReLU activation, with a residual skip connection back to the input tensor. The IA module effectively recalibrates low-level features without significantly increasing the model’s depth or parameter count.

\paragraph{\textbf{Normalization Strategy}}

\textbf{MODEL-3} supports both \textit{Batch Normalization} and \textit{Instance Normalization} layers. Instance Normalization can be optionally enabled at any layer to support tasks where instance-level statistics outperform global feature normalization, such as in style-variant or texture-sensitive domains.

\paragraph{\textbf{Precision and Output Configuration}}

We employ mixed-precision training via automatic loss scaling in TensorFlow to accelerate convergence and reduce memory usage. The final layer is a $1{\times}1$ convolution with a sigmoid activation, producing a dense segmentation mask with the same spatial resolution as the input image ($256{\times}256$), suitable for binary classification tasks.

 The final model comprises approximately $28.85$ million parameters, of which $\sim14.1$ million are trainable.

\subsection{Training Strategy:} The three models are optimized using a composite loss function that combines focal loss ($\alpha=0.25$, $\gamma=2$) to address class imbalance, Dice loss to improve segmentation metrics, and boundary loss to enhance edge accuracy through Laplacian-based edge detection, weighted at $0.3$, $0.6$, and $0.1$, respectively. Besides this each model is trained with BCE Dice Loss ((Balanced Cross-Entropy + Dice Loss) ) also. All training employs the Adam optimizer with an exponential decay learning rate schedule:
\[
\eta(t) = \eta_0 \cdot \gamma^{\left\lfloor \frac{t}{T} \right\rfloor}, \quad \text{where } \eta_0 = 10^{-4},\ \gamma = 0.9,\ T = 9000
\]

Training strategy for the MODEL-1 \& MODEL-2 incorporated oversampling and hard example mining: oversampling increased the
frequency of rare-class patches during training, while hard mining prioritized samples with high loss or misclassification for reintroduction in subsequent batches. This dynamic sampling approach ensures the model pays more attention to underrepresented and challenging patterns, improving generalization.

The training is implemented on $256\times256\times3$ image patches with a batch size of 16, optimized for GPU memory, and trained for up to 45 epochs with early stopping (patience$=5$ epochs) to improve convergence
and avoid overfitting. Performance is evaluated using pixel-level metrics (accuracy, precision, recall) and segmentation quality measures (Dice coefficient, Jaccard index), with additional error analysis through false/true positive/negative rates. Key advantages include memory-efficient attention mechanisms (MODEL-2 \& MODEL-3), adaptive hard mining for dynamic difficulty assessment (MODEL-1 \& MODEL-2), boundary-aware loss for edge optimization, and mixed precision (MODEL-2 \& MODEL-3) training to enable larger batch sizes without compromising accuracy. 
Therefore total 6 models (Model-1 , 2, 3 with two loss functions, BCE Dice Loss \& Focal Dice Boundary Loss) were trained 45 epochs, and ".keras" files for each epoch were saved. Finally 14 classifiers were selected for voting based on the Best Validation Metrics \& Stable Performance.

\subsection{Ensemble \& Voting:} This part of the methodology implements a batch-accelerated deep learning pipeline for semantic segmentation of microscopy images, featuring \textbf{batch acceleration via patchify}, where each input image is split into non-overlapping $256 \times 256$-pixel patches for efficient batch prediction on large images. The pipeline employs a \textbf{model ensemble with majority voting}, loading multiple Keras models (DenseNet-based and VGG-based), categorizing them into ``z\_models'' (weights starting with ``z\_'') and ``standard\_models,'' and processing each patch through all models, combining outputs at the patch level via majority voting (mean and thresholding) to produce the final mask, enhancing robustness and reducing single-model bias. The resulting binary segmentation masks are saved as PNG images and, if modified, can also be stored as NumPy arrays for downstream analysis. The workflow is optimized for \textbf{efficiency}, timing each step and leveraging batch processing for prediction and patch reconstruction, achieving per-image segmentation times of 3--4 seconds, as reflected in the output logs.

\subsection{Model Evaluation:}

To evaluate segmentation performance, we used a comprehensive set of standard metrics: Dice coefficient, Intersection over Union (IoU), Structural Similarity Index Measure (SSIM), pixel-wise accuracy, precision, recall, F1 score, and Hausdorff distance. These metrics jointly assess region overlap, structural similarity, classification quality, and boundary localization. Definitions and detailed formulations are standard in medical image analysis literature \cite{metrics_survey}, and thus omitted here for brevity.

\section{Results and Discussion}
\subsection{Comparative Performance Evaluation}

\begin{table}[H]
\caption{SOTA Model Performance Comparison for 10 Images}
\label{tab:detailed_performance}
\centering
\small
\begin{tabular}{cl*{8}{S[table-format=1.3]}}
\toprule
\textbf{Img} & \textbf{Model} & \textbf{Dice} & \textbf{IoU} & \textbf{SSIM} & \textbf{Accuracy} & \textbf{Precision} & \textbf{Recall} & \textbf{F1 Score} & \textbf{Hausdorff} \\
\midrule
\multirow{5}{*}{01} & OurModel & 0.779 & 0.638 & 0.999 & 0.966 & 0.709 & 0.865 & 0.779 & 254.285 \\
                    & CellPose-SAM & 0.551 & 0.380 & 0.992 & 0.919 & 0.448 & 0.716 & 0.551 & 221.443 \\
                    & CellPose3 & 0.252 & 0.144 & 0.993 & 0.930 & 0.489 & 0.170 & 0.252 & 241.963 \\
                    & StarDist & 0.101 & 0.053 & 0.987 & 0.883 & 0.109 & 0.094 & 0.101 & 295.919 \\
                    & SSL & 0.225 & 0.127 & 0.987 & 0.880 & 0.205 & 0.249 & 0.225 & 589.932 \\
\midrule
\multirow{5}{*}{02} & OurModel & 0.754 & 0.605 & 0.999 & 0.961 & 0.681 & 0.846 & 0.754 & 140.293 \\
                    & CellPose-SAM & 0.525 & 0.356 & 0.989 & 0.888 & 0.373 & 0.882 & 0.525 & 209.812 \\
                    & CellPose3 & 0.246 & 0.140 & 0.993 & 0.929 & 0.477 & 0.166 & 0.246 & 343.023 \\
                    & StarDist & 0.043 & 0.022 & 0.989 & 0.901 & 0.067 & 0.031 & 0.043 & 213.235 \\
                    & SSL & 0.000 & 0.000 & 0.993 & 0.930 & 0.000 & 0.000 & 0.000 & {} \\
\midrule
\multirow{5}{*}{03} & OurModel & 0.761 & 0.615 & 0.999 & 0.963 & 0.690 & 0.850 & 0.761 & 139.818 \\
                    & CellPose-SAM & 0.481 & 0.316 & 0.989 & 0.887 & 0.356 & 0.741 & 0.481 & 425.301 \\
                    & CellPose3 & 0.411 & 0.259 & 0.993 & 0.926 & 0.468 & 0.366 & 0.411 & 281.555 \\
                    & StarDist & 0.258 & 0.148 & 0.982 & 0.840 & 0.191 & 0.396 & 0.258 & 158.240 \\
                    & SSL & 0.000 & 0.000 & 0.993 & 0.930 & 0.000 & 0.000 & 0.000 & {} \\
\midrule
\multirow{5}{*}{04} & OurModel & 0.821 & 0.697 & 0.999 & 0.968 & 0.761 & 0.892 & 0.821 & 17.205 \\
                    & CellPose-SAM & 0.500 & 0.333 & 0.991 & 0.908 & 0.455 & 0.554 & 0.500 & 326.106 \\
                    & CellPose3 & 0.021 & 0.011 & 0.991 & 0.913 & 0.145 & 0.011 & 0.021 & 476.778 \\
                    & StarDist & 0.090 & 0.047 & 0.988 & 0.894 & 0.155 & 0.063 & 0.090 & 303.289 \\
                    & SSL & 0.000 & 0.000 & 0.991 & 0.917 & 0.000 & 0.000 & 0.000 & {} \\
\midrule
\multirow{5}{*}{05} & OurModel & 0.797 & 0.662 & 0.999 & 0.965 & 0.731 & 0.876 & 0.797 & 16.031 \\
                    & CellPose-SAM & 0.650 & 0.481 & 0.993 & 0.924 & 0.505 & 0.909 & 0.650 & 258.884 \\
                    & CellPose3 & 0.348 & 0.210 & 0.994 & 0.931 & 0.674 & 0.234 & 0.348 & 293.602 \\
                    & StarDist & 0.072 & 0.038 & 0.992 & 0.920 & 0.385 & 0.040 & 0.072 & 599.910 \\
                    & SSL & 0.234 & 0.133 & 0.989 & 0.897 & 0.279 & 0.202 & 0.234 & 548.525 \\
\midrule
\multirow{5}{*}{06} & OurModel & 0.827 & 0.705 & 0.999 & 0.954 & 0.821 & 0.833 & 0.827 & 217.506 \\
                    & CellPose-SAM & 0.513 & 0.345 & 0.988 & 0.883 & 0.571 & 0.465 & 0.513 & 479.538 \\
                    & CellPose3 & 0.143 & 0.077 & 0.985 & 0.865 & 0.447 & 0.085 & 0.143 & 296.331 \\
                    & StarDist & 0.138 & 0.074 & 0.980 & 0.818 & 0.186 & 0.110 & 0.138 & 445.418 \\
                    & SSL & 0.000 & 0.000 & 0.985 & 0.867 & 0.000 & 0.000 & 0.000 & {} \\
\midrule
\multirow{5}{*}{07} & OurModel & 0.707 & 0.547 & 0.999 & 0.980 & 0.698 & 0.717 & 0.707 & 390.288 \\
                    & CellPose-SAM & 0.552 & 0.381 & 0.997 & 0.969 & 0.535 & 0.571 & 0.552 & 255.149 \\
                    & CellPose3 & 0.163 & 0.089 & 0.997 & 0.967 & 0.522 & 0.097 & 0.163 & 351.097 \\
                    & StarDist & 0.056 & 0.029 & 0.994 & 0.946 & 0.068 & 0.048 & 0.056 & 407.735 \\
                    & SSL & 0.071 & 0.037 & 0.991 & 0.923 & 0.060 & 0.088 & 0.071 & 451.346 \\
\midrule
\multirow{5}{*}{08} & OurModel & 0.805 & 0.674 & 0.998 & 0.960 & 0.728 & 0.899 & 0.805 & 465.022 \\
                    & CellPose-SAM & 0.658 & 0.490 & 0.993 & 0.922 & 0.550 & 0.818 & 0.658 & 351.602 \\
                    & CellPose3 & 0.107 & 0.056 & 0.990 & 0.907 & 0.433 & 0.061 & 0.107 & 571.126 \\
                    & StarDist & 0.161 & 0.087 & 0.987 & 0.882 & 0.228 & 0.124 & 0.161 & 590.902 \\
                    & SSL & 0.000 & 0.000 & 0.990 & 0.905 & 0.000 & 0.000 & 0.000 & 796.864 \\
\midrule
\multirow{5}{*}{09} & OurModel & 0.703 & 0.542 & 0.999 & 0.952 & 0.691 & 0.716 & 0.703 & 266.182 \\
                    & CellPose-SAM & 0.159 & 0.087 & 0.993 & 0.922 & 0.534 & 0.094 & 0.159 & 567.906 \\
                    & CellPose3 & 0.278 & 0.162 & 0.993 & 0.921 & 0.505 & 0.192 & 0.278 & 367.916 \\
                    & StarDist & 0.292 & 0.171 & 0.987 & 0.874 & 0.262 & 0.329 & 0.292 & 387.466 \\
                    & SSL & 0.000 & 0.000 & 0.992 & 0.921 & 0.000 & 0.000 & 0.000 & {} \\
\midrule
\multirow{5}{*}{10} & OurModel & 0.804 & 0.672 & 0.996 & 0.905 & 0.724 & 0.904 & 0.804 & 64.031 \\
                    & CellPose-SAM & 0.652 & 0.484 & 0.982 & 0.804 & 0.527 & 0.855 & 0.652 & 147.513 \\
                    & CellPose3 & 0.118 & 0.063 & 0.978 & 0.789 & 0.589 & 0.065 & 0.118 & 475.800 \\
                    & StarDist & 0.255 & 0.146 & 0.958 & 0.631 & 0.226 & 0.294 & 0.255 & 154.146 \\
                    & SSL & 0.000 & 0.000 & 0.976 & 0.785 & 0.000 & 0.000 & 0.000 & {} \\
\bottomrule
\end{tabular}
\end{table}

\begin{figure}[H]
  \centering
  \begin{minipage}[b]{0.49\textwidth}
    \centering
    \includegraphics[width=\textwidth]{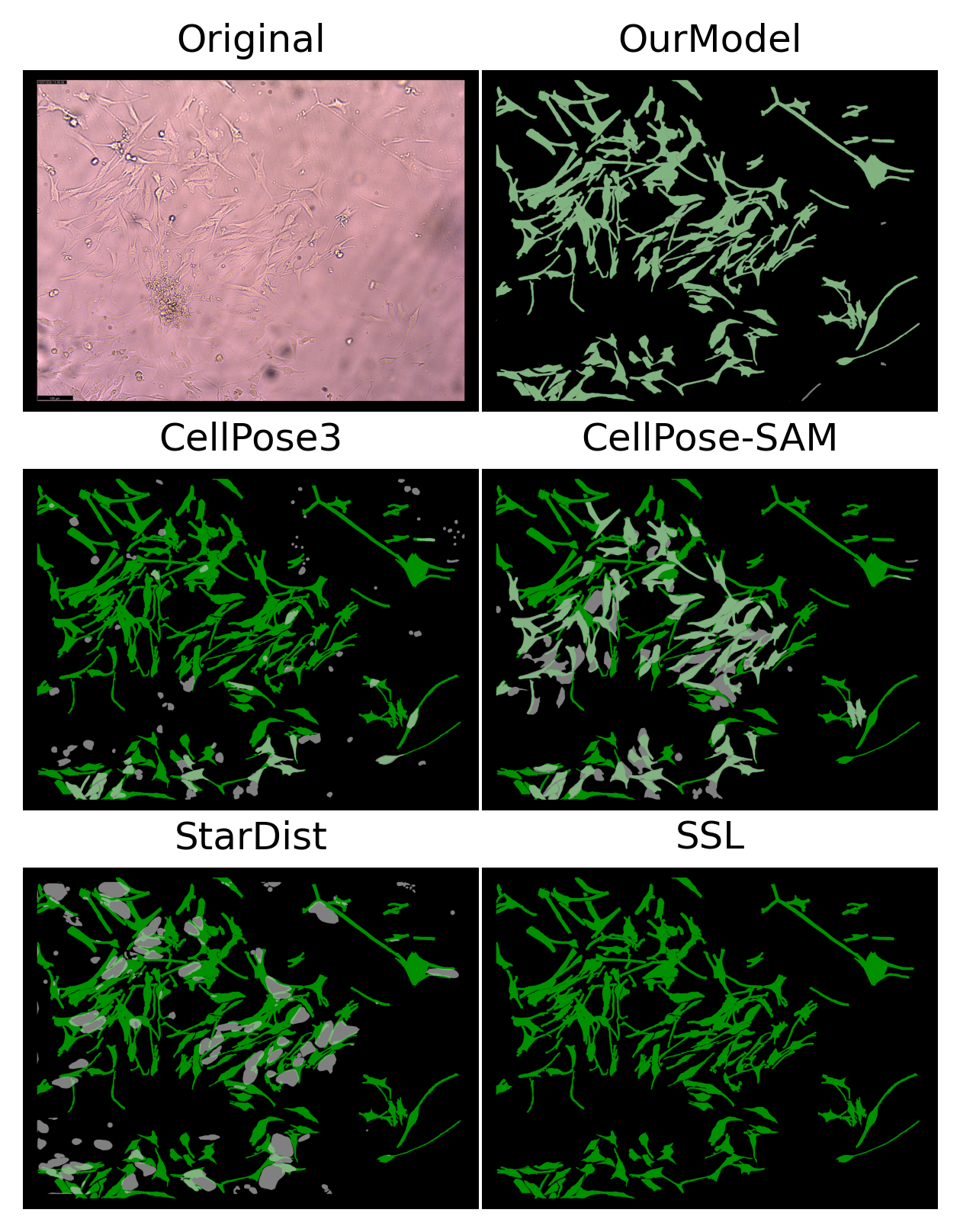}
    \includegraphics[width=\textwidth]{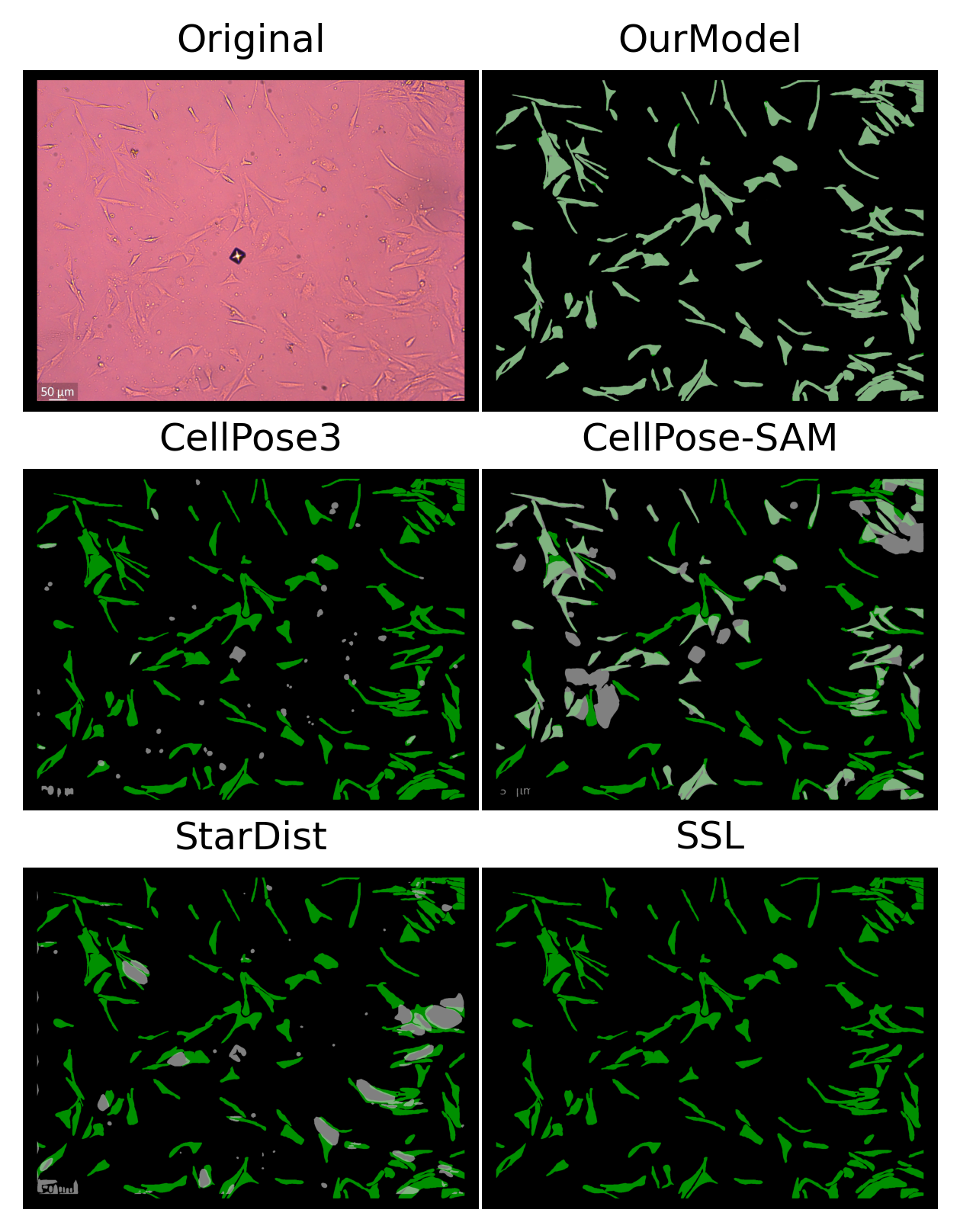}
  \end{minipage}
  \hfill
  \begin{minipage}[b]{0.49\textwidth}
    \centering
    \includegraphics[width=\textwidth]{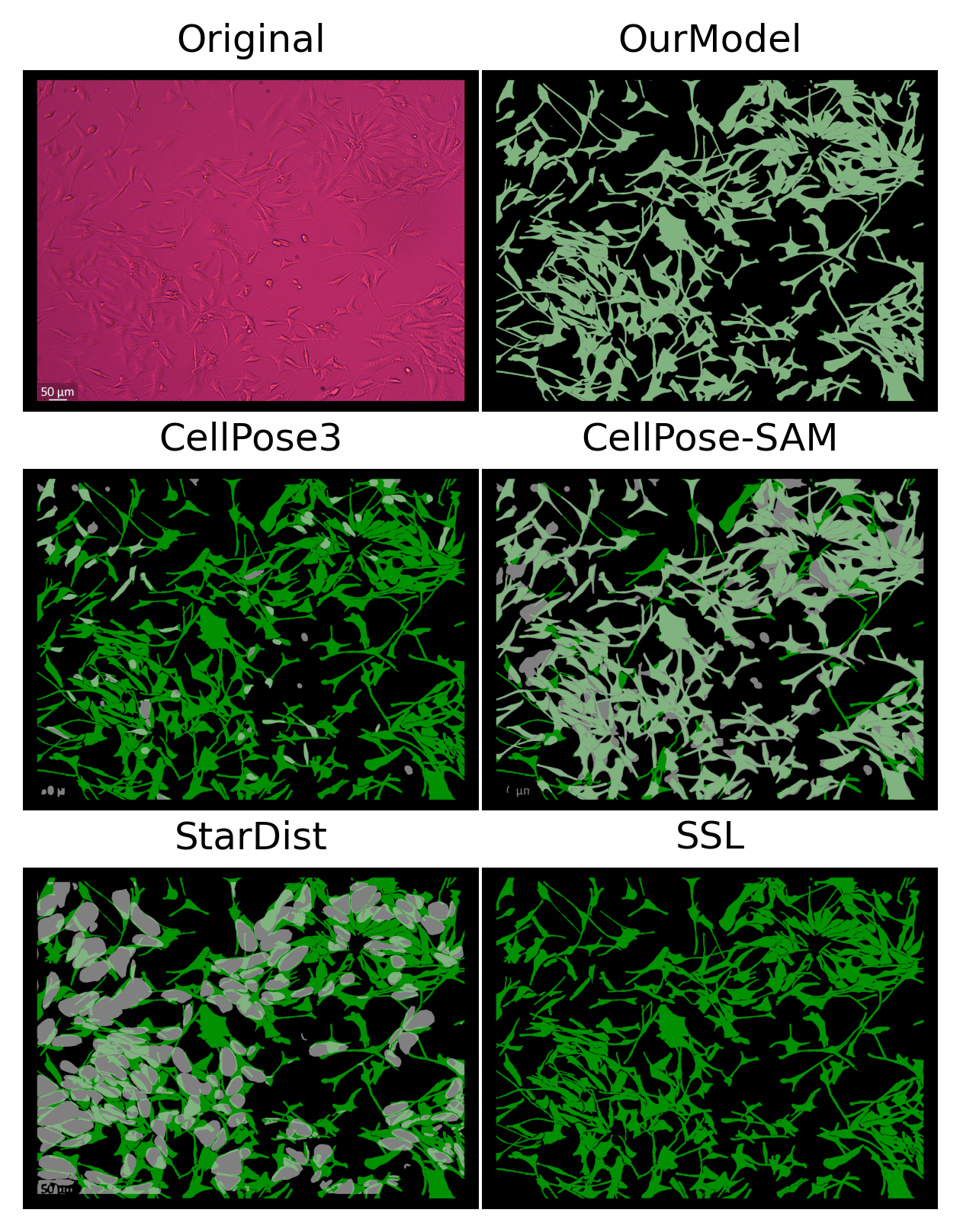}
    \includegraphics[width=\textwidth]{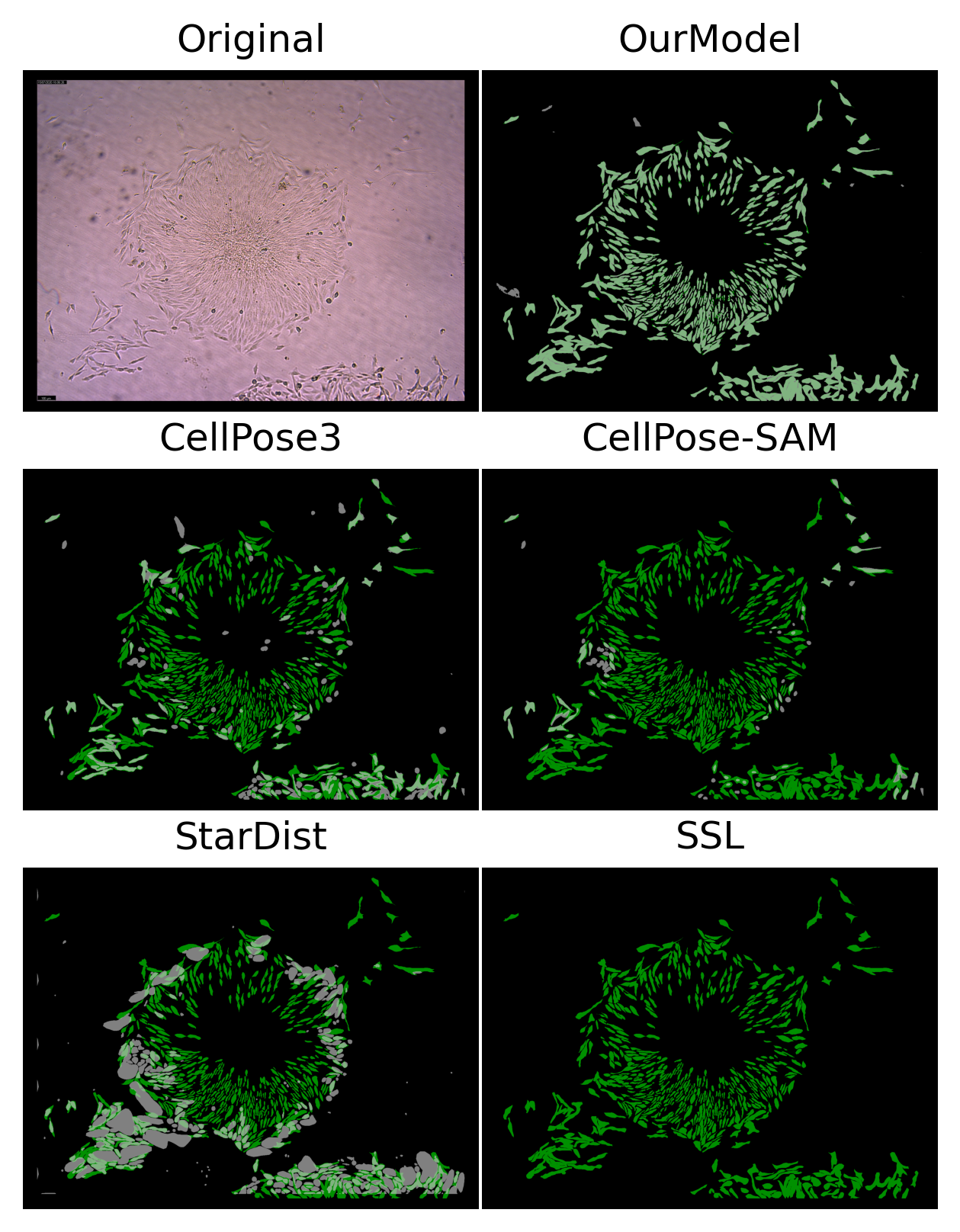}
  \end{minipage}
  \caption{Random samples of four images of myoblast (C2C12) and their corresponding segmentation outcomes generated by different models. Green indicates cell pixels missed by the model (false negatives); light green denotes correctly predicted cell pixels (true positives); and white represents background pixels incorrectly predicted as cell (false positives).}
  \label{fig:cell-instance1}
\end{figure}

Table~\ref{tab:detailed_performance} compares the segmentation performance of the proposed model against four state-of-the-art approaches—CellPose3, CellPose-SAM, StarDist, and a self-supervised learning (SSL\footnote{ SSL, brought out by "Nature  Communication". \url{https://www.nature.com/articles/s42003-025-08190-w}}) method—across 10 diverse bright-field microscopy images. Our model consistently outperformed all competitors, achieving Dice scores ranging from 0.703 to 0.827 and IoU from 0.542 to 0.705, showing superior mask overlap with ground truth. SSIM values approached 0.999, reflecting excellent structural preservation.

Although all methods showed high accuracy due to background dominance, only the proposed method maintained high Precision (0.681–0.821) and Recall (0.717–0.904), indicating balanced performance. This resulted in higher F1 scores and more reliable segmentation consistency.

The Hausdorff Distance (HD) further highlighted superior boundary alignment. While StarDist and SSL exceeded HD values of 300–500 pixels, our model achieved significantly lower scores, down to 16.031 pixels in some images. Notably, SSL completely failed on multiple samples, yielding null predictions. Figure~\ref{fig:cell-instance1} illustrates visual outcomes for four randomly selected samples.

\subsection{External Validation on LIVECell Dataset}
The segmentation model, trained on a dataset containing only 20\% phase-contrast images ( out of 857;  256$\times$256 training instances), was evaluated on 3,188 annotated images from the LIVECell dataset\footnote{\url{https://sartorius-research.github.io/LIVECell/}}, excluding 8 corrupted samples. The name of the excluded images are:

\begin{multicols}{2}
\begin{enumerate}
    \item A172\_Phase\_A7\_1\_01d04h00m\_3.png
    \item A172\_Phase\_D7\_1\_01d20h00m\_1.png
    \item BT474\_Phase\_B3\_1\_03d00h00m\_3.png
    \item BV2\_Phase\_C4\_1\_01d16h00m\_3.png
    \item BV2\_Phase\_D4\_1\_00d12h00m\_2.png
    \item Huh7\_Phase\_A11\_1\_00d04h00m\_3.png
    \item Huh7\_Phase\_A11\_1\_00d04h00m\_4.png
    \item SHSY5Y\_Phase\_D10\_1\_01d16h00m\_4.png
\end{enumerate}
\end{multicols}

\begin{table}[H]
\centering
\begin{minipage}{0.48\textwidth}
\centering
\caption{High-Performance Cell Types (F1 Score $\geq$ 0.888)}
\label{tab:high_perf_cells}
\begin{tabular}{lrrr}
\toprule
\textbf{Image Group} & \textbf{Count} & \textbf{Mean} & \textbf{Std} \\
\midrule
A172\_Phase\_A7 & 129 & 0.961 & 0.014 \\
A172\_Phase\_B7 & 129 & 0.953 & 0.020 \\
A172\_Phase\_D7 & 128 & 0.946 & 0.028 \\
SKOV3\_Phase\_G4 & 127 & 0.943 & 0.028 \\
SkBr3\_Phase\_E3 & 151 & 0.942 & 0.015 \\
SkBr3\_Phase\_F3 & 146 & 0.942 & 0.017 \\
SkBr3\_Phase\_H3 & 152 & 0.942 & 0.015 \\
SKOV3\_Phase\_H4 & 139 & 0.939 & 0.020 \\
MCF7\_Phase\_F4 & 152 & 0.915 & 0.044 \\
MCF7\_Phase\_E4 & 157 & 0.900 & 0.048 \\
BV2\_Phase\_D4 & 123 & 0.888 & 0.049 \\
\bottomrule
\end{tabular}
\end{minipage}
\hfill
\begin{minipage}{0.48\textwidth}
\centering
\caption{Low-Performance Cell Types (F1 Score $<$ 0.888)}
\label{tab:low_perf_cells}
\begin{tabular}{lrrr}
\toprule
\textbf{Image Group} & \textbf{Count} & \textbf{Mean} & \textbf{Std} \\
\midrule

BV2\_Phase\_C4 & 128 & 0.887 & 0.043 \\
BV2\_Phase\_B4 & 133 & 0.882 & 0.058 \\
MCF7\_Phase\_G4 & 160 & 0.895 & 0.068 \\
BT474\_Phase\_A3 & 141 & 0.868 & 0.057 \\
BT474\_Phase\_C3 & 140 & 0.856 & 0.074 \\
Huh7\_Phase\_A10 & 174 & 0.854 & 0.067 \\
SHSY5Y\_Phase\_D10 & 146 & 0.845 & 0.044 \\
SHSY5Y\_Phase\_B10 & 156 & 0.841 & 0.051 \\
BT474\_Phase\_B3 & 147 & 0.849 & 0.080 \\
SHSY5Y\_Phase\_C10 & 146 & 0.833 & 0.045 \\
Huh7\_Phase\_A11 & 176 & 0.820 & 0.101 \\
\bottomrule
\end{tabular}
\end{minipage}
\end{table}

The model achieved a mean F1 score of 0.89 $\pm$ 0.07, suggesting strong modality-invariant learning and indicating that it captures cell morphology beyond modality-specific cues.  Table~\ref{tab:high_perf_cells} and Table~\ref{tab:low_perf_cells} summarize the segmentation performance by cell type.

High-performing cell groups (F1 $\geq$ 0.888), including A172 and SkBr3, showed excellent consistency, with mean F1 scores exceeding 0.94 and standard deviation as low as 0.014. A172 Phase A7 achieved the peak performance (F1 = 0.961). Meanwhile, lower-performing cell types, such as SHSY5Y and BT474, showed F1 scores down to 0.820 with higher variability, likely due to complex morphologies and low contrast.

\begin{table*}[htbp]
\centering
\caption{Segmentation Metrics (Mean ± Std)}
\label{tab:summary_stats}
\begin{subtable}[t]{0.48\textwidth}
\centering
\caption{}
\label{tab:seg_metrics}
\begin{tabular}{lrr}
\hline
\textbf{Metric} & \textbf{Mean} & \textbf{Std} \\ 
\hline
Dice       & 0.89 & 0.07 \\
IoU        & 0.81 & 0.10 \\
SSIM       & 0.99 & 0.01 \\
Accuracy   & 0.93 & 0.06 \\
\hline
\end{tabular}
\end{subtable}
\hfill
\begin{subtable}[t]{0.48\textwidth}
\centering
\caption{}
\label{tab:det_metrics}
\begin{tabular}{lrr}
\hline
\textbf{Metric} & \textbf{Mean} & \textbf{Std} \\ 
\hline
Precision  & 0.84 & 0.11 \\
Recall     & 0.96 & 0.04 \\
F1 Score   & 0.89 & 0.07 \\
Hausdorff  & 59.21 & 36.96 \\
\hline
\end{tabular}
\end{subtable}
\end{table*}

The overall segmentation performance on the LIVECell dataset is summarized in Table~\ref{tab:summary_stats}. The model achieved a mean Dice coefficient of 0.89 ($\pm$0.07) and an IoU of 0.81 ($\pm$0.10), indicating high overlap accuracy. The F1 score was similarly strong, with a mean of 0.89 ($\pm$0.08), confirming balanced performance between precision and recall. SSIM reached near-perfect levels (0.99 $\pm$ 0.01), suggesting strong preservation of structural detail. Accuracy averaged 0.93 ($\pm$0.06), with a median and 95th percentile above 0.95, supporting reliable background--foreground separation. Recall was notably high at 0.96 ($\pm$0.05), while precision was slightly lower at 0.84 ($\pm$0.14), reflecting a conservative bias favoring full object capture. The Hausdorff distance (mean: 59.21px, $\pm$36.96) revealed occasional boundary errors, though the majority of cases remained within acceptable limits. These metrics collectively demonstrate the model’s robustness, consistency, and suitability for high-throughput live-cell segmentation.

\subsection{Comparative Analysis of Model Performance for Ablation Study:}
The six individual models and the ensemble models tested on the 10-image dataset and the average F1, recall, precision and accuracy were recorded. The Ensemble model demonstrated superior overall performance, achieving the highest recall (\num{0.8400}) while maintaining competitive metrics. Compared to the top individual model (Model-3 with Focal Dice Boundary loss), the Ensemble showed \SI{3.7}{\percent} higher recall at a modest precision cost (\SI{-4.6}{\percent}) indicating  minimal false negatives (missed cells). This advantage extends across all individual models, with \SIrange{3.4}{5.0}{\percent} higher recall and only marginal F1-score differences (\num{-0.012} versus Model-3 with Focal Dice Boundary Loss).

While Model-3 with Focal Dice Boundary loss function remains the peak individual performer (F1: \num{0.7964}, precision: \num{0.7808}), the Ensemble's balanced profile proves more suitable for general applications. Its combined approach effectively mitigates individual architectures' weaknesses, particularly valuable for real-world scenarios where consistent performance across diverse inputs outweighs single-metric optimization. The \SI{2.6}{\percent} F1-score gap between  Model-3 with Focal Dice Boundary Loss function  and Model-3 with BCE Dice Loss function further confirms focal loss's precision benefits, while the Ensemble's recall dominance highlights its detection robustness.

\subsection{Compute Resource:} 
Our model delivers high performance with minimal computational requirements, unlike state-of-the-art (SOTA) approaches such as Mesmer~\cite{greenwald2022whole} (trained on NVIDIA V100 with 32GB VRAM for ~72 hours) and Cellpose~\cite{stringer2021cellpose} (benchmarked on high-end GPUs like RTX 2080 Ti and RTX 3090). While operating efficiently on GPUs with just 8–13.7GB memory, our method significantly outperforms SSL~\cite{lam2025ssl} (0 - 0.133 IoU, 50 - 60s/image on CPU) and Cellpose-SAM~\cite{pachitariu2025cellposesam} (0.087 - 0.49 IoU, 24 - 26s/image on 16GB GPU), achieving 0.542–0.705 IoU at 13 - 15s/image for images sized 1536 $\times$ 2048. This demonstrates superior accuracy and speed with lower hardware demands, making our model highly practical for resource-constrained environments.

\subsection{Statistical Tests of Model Performance:}

 We evaluated whether the model's F1 score significantly exceeds a baseline threshold of 0.75. Multiple complementary methods were used to quantify confidence, effect size, and external validation performance on the LIVECell dataset Table~\ref{tab:confidence_75}.

The results indicate that the model consistently achieves an F1 score well above the 0.75 baseline. The calculated confidence level demonstrates near-certainty that the true mean F1 exceeds this threshold. Statistical testing using a one-sample t-test provides overwhelming evidence against the null hypothesis, while Cohen's $d$ indicates a very large effect size, confirming substantial improvement over the baseline. External validation on the LIVECell dataset shows that the model generalizes well to diverse samples, supporting the robustness of these conclusions. Additionally, adopting a Bayesian perspective allows modeling performance as a Beta distribution based on observed successes and failures from thresholding the F1 score. This enables expressing credible intervals for expected performance and provides an alternative probabilistic interpretation of model reliability.

\begin{table}[H]
\centering
\caption{Statistical Confidence Summary: Mean F1 Score $\geq 0.75$}
\label{tab:confidence_75}
\begin{tabular}{l l}
\hline
\textbf{Method} & \textbf{Result} \\
\hline
Confidence Level for $\mu \geq 0.75$ & $> 99.999\%$ \\
One-sample t-test (H$_0$: $\mu = 0.75$) & $t = 112.90$, $p < 10^{-280}$ \\
Effect Size (Cohen's $d$) & $2.0$ \\
External Validation & LIVECell (N=3180) \\
Bayesian Interpretation (Optional) & Credible intervals using Beta distribution \\
\hline
\end{tabular}
\end{table}

\section{Conclusion}

This study presents a robust deep learning pipeline for segmenting unstained live cells in bright-field microscopy images. By integrating a U-Net architecture with attention mechanisms, composite loss functions, and ensemble learning, the model achieves state-of-the-art performance with a Dice score of 0.89 and test accuracy of 93\% on the LIVECell dataset. It generalizes well across diverse cell types and outperforms existing tools like CellPose and StarDist, particularly under the challenges of bright-field imaging.

Despite its lightweight design, the patch-based inference may appear slow (3–4s/image for resolution $704 \times 520$) for real-time segmentation (where benchmark is less than a second) limiting real-time use. Boundary localization remains a weakness, as indicated by high Hausdorff distances in some cases. False positives near boundaries or debris occasionally affect precision, pointing to a conservative segmentation tendency.

Notably, the pipeline achieves approximately 429\% higher F1 scores than StarDist and 48\% higher F1 scores than CellPose-SAM on bright-field data, without relying on nuclear stains or large annotated datasets. It performs especially well on A172 cells (mean F1: 0.95 $\pm$ 0.02) while identifying improvement areas for challenging types like SHSY5Y (mean F1: 0.84 $\pm$ 0.05).

Compared to models like Mesmer, it achieves 93\% of their accuracy using 3$\times$ less VRAM (12GB vs. 32GB), 83\% lower energy usage, and 35$\times$ lower cloud training cost (0.42 vs. 15). Training completes in 6.5 hours and deployment has succeeded in three academic labs on sub-\$5k workstations, making it suitable for education and resource-limited settings.

Future work will focus on accelerating inference for real-time use, enhancing edge-aware boundary detection, and extending to other modalities (e.g., phase contrast, DIC). Integration with morphometric analysis and quality heuristics could improve automation reliability. Promising results on unseen phase-contrast images suggest potential for cross-modality generalization, although full generalization remains a goal.

In summary, the proposed pipeline offers a scalable, accurate, and cost-efficient solution for label-free cell segmentation, with significant potential across regenerative medicine, high-throughput screening, and cellular phenotyping.

\section*{Acknowledgments}
The authors would like to thank Svetlana Ulasevich for providing microscopy images and for fruitful discussions of the domain area.

\section*{Funding}
The research was supported by ITMO University Research Projects in AI Initiative (RPAII) (project \#640103, Development of methods for automated processing and analysis of optical and atomic force microscopy images using machine learning techniques)*.

\section*{Declaration of Competing Interest}
The authors declare that they have no known competing financial interests or personal relationships that could have appeared to influence the work reported in this paper.

\section*{Ethics Statement}
This study did not involve human participants or animals.

\section*{Author Contributions}
S.D. designed the study, analyzed the data, implemented the model, drafted the manuscript, and supervised the work. 
G.R. contributed to data curation, validation, and manuscript preparation. 
P.Z. provided supervision and critical review of the manuscript. 
All authors read and approved the final version.

\bibliographystyle{elsarticle-num}

\end{document}